\documentclass[aps,prl,twocolumn,superscriptaddress,showpacs]{revtex4}

\usepackage[dvips]{graphicx}
\usepackage{xspace}

\begin{document}

% definitions here
\newcommand{\superk}    {Super-Kamiokande\xspace}       
\newcommand{\nux}       {\ensuremath{\nu_{x}}\xspace}
\newcommand{\nue}       {\ensuremath{\nu_{e}}\xspace}
\newcommand{\nuebar}    {\ensuremath{\overline{\nu}_e}\xspace}
\newcommand{\numu}      {\ensuremath{\nu_{\mu}}\xspace}
\newcommand{\numubar}   {\ensuremath{\overline{\nu}_\mu}\xspace}
\newcommand{\nutau}     {\ensuremath{\nu_{\tau}}\xspace}
\newcommand{\nusterile} {\ensuremath{\nu_{sterile}}\xspace}
\newcommand{\mue}       {\ensuremath{\nu_\mu \rightarrow \nu_{e}}\xspace}
\newcommand{\mumu}      {\ensuremath{\nu_\mu \rightarrow \nu_\mu}\xspace}
\newcommand{\mutau}     {\ensuremath{\nu_\mu \rightarrow \nu_{\tau}}\xspace}
\newcommand{\musterile} {\ensuremath{\nu_\mu \rightarrow \nu_{sterile}}\xspace}
\newcommand{\dms}       {\ensuremath{\Delta m^2}\xspace}
\newcommand{\dmsatm}    {\ensuremath{\Delta m^2_{atm}}\xspace}
\newcommand{\dmsmue}    {\ensuremath{\Delta m_{\mu{}e}^2}\xspace}
\newcommand{\dmsmumu}   {\ensuremath{\Delta m_{\mu{}\mu{}}^2}\xspace}
\newcommand{\dmsonethree}     {\ensuremath{\Delta m_{13}^2}\xspace}
\newcommand{\sstt}      {\ensuremath{\sin^2 2 \theta}\xspace}
\newcommand{\ssttmue}   {\ensuremath{\sin^2 2 \theta_{\mu{}e}}\xspace}
\newcommand{\thetamue}  {\ensuremath{\theta_{\mu{}e}}\xspace}
\newcommand{\thetamumu} {\ensuremath{\theta_{\mu{}\mu{}}}\xspace}
\newcommand{\Rnqe}      {\ensuremath{R_{nqe}}\xspace}

\title{Search for Electron Neutrino Appearance in
  a 250 km Long-baseline Experiment}

%%% Authors %%%
% Affiliation definitions and then the authors
\newcommand{\Osaka}{\affiliation{ Department of Physics, Osaka
    University, Toyonaka, Osaka 560-0043, JAPAN}}

\newcommand{\SNU}{\affiliation{ Department of Physics, Seoul National
    University, Seoul 151-742, KOREA}}

\newcommand{\Kobe}{\affiliation{ Kobe University, Kobe, Hyogo
    657-8501, JAPAN}}

\newcommand{\UW}{\affiliation{ Department of Physics, University of
    Washington, Seattle, WA 98195-1560, USA }}

\newcommand{\UCI}{\affiliation{ Department of Physics and Astronomy,
    University of California, Irvine, Irvine, CA 92697-4575, USA }}

\newcommand{\CNU}{\affiliation{ Department of Physics, Chonnam
    National University, Kwangju 500-757, KOREA}}

\newcommand{\ICRR}{\affiliation{ Institute for Cosmic Ray Research,
    University of Tokyo, Kashiwa, Chiba 277-8582, JAPAN}}

\newcommand{\Kyoto}{\affiliation{ Department of Physics, Kyoto
    University, Kyoto 606-8502, JAPAN}}

\newcommand{\Tohoku}{\affiliation{ Research Center for Neutrino
    Science, Tohoku University, Sendai, Miyagi 980-8578, JAPAN}}

\newcommand{\KEK}{\affiliation{ High Energy Accelerator Research Organization (KEK), Tsukuba, Ibaraki 305-0801, JAPAN }}

\newcommand{\SUNY}{\affiliation{ Department of Physics and Astronomy,
    State University of New York, Stony Brook, NY 11794-3800, USA}}

\newcommand{\Okayama}{\affiliation{ Department of Physics, Okayama
    University, Okayama, Okayama 700-8530, JAPAN}}

\newcommand{\BU}{\affiliation{ Department of Physics, Boston
    University, Boston, MA 02215, USA}}

\newcommand{\Hawaii}{\affiliation{ Department of Physics and
    Astronomy, University of Hawaii, Honolulu, HI 96822, USA}}

\newcommand{\Warsaw}{\affiliation{ Institute of Experimental Physics,
    Warsaw University, 00-681 Warsaw, POLAND }}

\newcommand{\SINS}{\affiliation{ A. Soltan Institute for Nuclear
    Studies, 00-681 Warsaw, POLAND}}

\newcommand{\KU}{\affiliation{ Department of Physics, Korea
    University, Seoul 136-701, KOREA}}

\newcommand{\Niigata}{\affiliation{ Department of Physics, Niigata
    University, Niigata, Niigata 950-2181, JAPAN}}

\newcommand{\Dongshin}{\affiliation{ Department of Physics, Dongshin
    University, Naju 520-714, KOREA}}

\newcommand{\MIT}{\affiliation{ Department of Physics, Massachusetts
    Institute of Technology, Cambridge, MA 02139, USA}}

\newcommand{\TSU}{\affiliation{ Department of Physics, Tokyo
    University of Science, Noda, Chiba 278-0022, JAPAN}}

\newcommand{\Miyakyo}{\affiliation{ Department of
    Physics, Miyagi University of Education, Sendai 980-0845, JAPAN}}

% The alternative affilations will show up in the references

\newcommand{\now}{\altaffiliation{For current affiliations see
    http://neutrino.kek.jp/present-addresses0401.ps}}

% not used for now
\newcommand{\ICEPP}{\altaffiliation{ International Center for Elementary Particle Physics,
    University of Tokyo, Tokyo, 113-0033, JAPAN}}

\newcommand{\KRISS}{\altaffiliation{ Korea Research Institute of Standard and Science, Yuseong, Daejeon, 305-600, KOREA}}

\newcommand{\Tokai}{\altaffiliation{ Department of Physics, Tokai
    University, Hiratsuka, Kanagawa 259-1292, JAPAN}}

\newcommand{\UPnow}{\altaffiliation{ Present address: University of
  Pittsburgh, Pittsburgh, PA 15260, USA}}

\newcommand{\Hillnow}{\altaffiliation{ Present address: California State
    University, Dominghez Hills, USA}}

\newcommand{\Jangnow}{\altaffiliation{ Present address: Seokang College,
    Kwangju, 500-742, KOREA}}

\newcommand{\Kainow}{\altaffiliation{ Present address: Department of
    Physics, University of Utah, Salt Lake City, UT 84112, USA}}

\newcommand{\Marunow}{\altaffiliation{ Present address: The Enrico Fermi
    Institute, University of Chicago, Chicago, IL 60637, USA}}

\newcommand{\Maugernow}{\altaffiliation{ Present address: California
    Institute of Technology, California 91125, USA}}

\newcommand{\Nagoyanow}{\altaffiliation{ Present address: Department of
    Physics, Nagoya University, Nagoya, Aichi 464-8602, JAPAN}}

\newcommand{\DankaSup}{\altaffiliation{ Supported by the Polish Committee
    for Scientific Research}}

\author{M.H.Ahn}\SNU
\author{S.Aoki}\Kobe
\author{Y.Ashie}\ICRR
\author{H.Bhang}\SNU
\author{S.Boyd}\now\UW
\author{D.Casper}\UCI
\author{J.H.Choi}\CNU
\author{S.Fukuda}\ICRR
\author{Y.Fukuda}\Miyakyo
%\author{W.Gajewski}\UCI
\author{R.Gran}\UW
% ...

\collaboration{The K2K Collaboration}\noaffiliation
%%%

\date{\today}

%%% Abstract %%%
\begin{abstract}
\label{chp:abstract}

We present a search for electron neutrino appearance from accelerator
produced muon neutrinos in the K2K long baseline neutrino experiment.  One
candidate event is found in the data corresponding to an exposure of
$4.8\times{}10^{19}$ protons on target.  The expected background in the
absence of neutrino oscillations is estimated to be $2.4\pm{}0.6$ events
and is dominated by mis-identification of events from neutral current
$\pi^0$ production.  We exclude the \numu to \nue oscillations at 90\% C.L.
for the effective mixing angle in 2-flavor approximation of
$\ssttmue{} (\simeq \frac{1}{2} \sin^2 2 \theta_{13} ) > 0.15$ at $\dmsmue{} = 2.8\times 10^{-3}$eV$^2$, the best
fit value of the \numu disappearance analysis in K2K.  The most stringent
limit of $\ssttmue{} < 0.09$ is obtained at $\dmsmue{} =
6\times{}10^{-3}$eV$^2$.

\end{abstract}
%%%

\pacs{PACS numbers: 14.60.Pq, 13.15.+g, 23.40.Bw, 95.55.Vj}

\maketitle

%%% Introduction %%%
In 1998, the Super-Kamiokande (SK) collaboration reported evidence of
neutrino oscillation based on atmospheric neutrino observations favoring
large mixing between \numu{} and \nutau{} and a \dms{} near $2.2 \times
10^{-3}\mbox{eV}^2$~\cite{Fukuda:1998mi}.  Subsequently, solar neutrino
data from various experiments have indicated $\nu_e$ disappearance as a
result of neutrino oscillations to other active neutrino flavors (\numu{}
or \nutau{}) with large mixing and a $\Delta m^2$ near $5 \times
10^{-5}\mbox{eV}^2$~\cite{Fukuda:2002ds,Ahmad:2002de}.  The KamLAND
experiment also observes a deficit of reactor \nuebar consistent with the
same parameter values~\cite{KamLAND:2003} as those in the solar neutrinos.
Recently, the KEK to Kamioka long-baseline neutrino oscillation
experiment~(K2K)~\cite{K2Kosci:2002up} reported indications of $\numu
\rightarrow \nux$ oscillation using an accelerator produced \numu{} beam.
The measurement of \numu{} disappearance in K2K results in neutrino
oscillation parameters which are consistent with the values derived from
the atmospheric neutrino oscillations.

Measurements of atmospheric and solar neutrinos suggest mixing between all
neutrino flavors.  The \nue appearance is predicted at the same \dms{} as
the one measured using the atmospheric neutrinos (\dmsatm{}) in the
framework of 3-flavor neutrino oscillations with certain parameter values.
%The probability of \nue appearance in the manner of 2-flavor approximation
The probability of \nue appearance in the 2-flavor approximation
is expressed by
\begin{equation}
  P(\mue{}) 
  = \sin^2 2\thetamue{}  
  \sin^2(1.27\dmsmue{} L/E),
  \label{eq:appearance}
\end{equation}
where \thetamue{} and \dmsmue{} are the effective mixing angle and mass
squared difference between the mass eigenstates involved, respectively, for
\nue{} appearance; $L$ is neutrino path length in kilometers; and $E$ is
the neutrino energy in GeV.
In 3-flavor framework with ``one mass scale dominance''~\cite{Fogli:1995uu},
the effective mixing parameter is related to the mixing angle $\theta_{13}$ by
\begin{equation}
  \sin^2 2\thetamue{} \simeq \sin^2 \theta_{23} \sin^2 2 \theta_{13}
  \simeq \frac{1}{2} \sin^2 2 \theta_{13}, (\theta_{23} \simeq \pi/4).
\label{eq:ssttmue-2313}
\end{equation}
%Other descriptions of the relation can be found in Ref.~\cite{PDBook}.
In this letter, we report results from the first search for \nue{}
appearance using \numu{} beam in the K2K experiment sensitive to the
\dmsatm{} region.
%%%

%%% Overview %%%
In K2K, an almost pure ($98$\%) \numu{} beam with a mean energy of 1.3~GeV
is produced with the KEK proton synchrotron (KEK-PS)~\cite{k2kdetect00}.
The fraction of \nue{} is approximately $1$\% and the remainder are
\numubar{}.  Twelve GeV protons from the KEK-PS hit an aluminum target
embedded inside a pulsed magnetic horn system which focuses positively
charged secondary particles, mainly pions, toward a far detector located
250~km far from KEK.  The secondary particles decay to produce a neutrino
beam.  The stability of the pulse-by-pulse beam direction is checked by
monitoring muons from pion decay with a set of ionization chambers and
silicon pad detectors following the beam dump.  The measurements from these
monitors show that the beam is directed to within 1~mrad of the far
detector, SK~\cite{SKNIM:2003}, which is a 50~kt Water Cherenkov detector
located in Kamioka, Gifu Prefecture in Japan.
 
To reject cosmic-ray and atmospheric neutrino background, the global
positioning system is employed at both the KEK and SK sites to synchronize
between beam spills and events observed in SK~\cite{GPS}.  The neutrino
flux at KEK is measured by a near detector complex consisting of a
1~kt water Cherenkov detector (1KT) and a fine-grained detector (FGD)
system. The FGD consists of a scintillating fiber detector with segmented
water targets (SciFi)~\cite{Suzuki:2000nj}, a plastic scintillator
hodoscope (PSH), a lead-glass calorimeter~(LG), and a muon range
detector~(MRD)~\cite{Ishii:2002nj}.  The \numu flux at SK is estimated by
extrapolating the measured flux at KEK using predicted flux ratio between
SK and KEK (far/near ratio).  The far/near ratio is evaluated by a beam
Monte Carlo (MC) simulation as a function of neutrino energy and is
validated using secondary pion kinematic distributions measured with a gas
Cherenkov detector~\cite{Maruyama} downstream of the horn system.

Electron neutrino events in SK are selected assuming \nue charged current
quasi-elastic (CC-QE) interactions in an oxygen nucleus, i.e. $\nu_e + n
\to e + p$.  Since the proton momentum in the reaction is typically below
Cherenkov threshold in water, only the electron is visible. Thus, a single
electron-like Cherenkov ring is the signature of $\nu_e$ appearance.  The
\nue{} contamination in the beam is estimated by the beam MC simulation
which predicts the ratio of the number of \nue interactions to that of
\numu to be 0.9\% in SK.  This estimate is checked by a measurement of the
\nue{} fraction at KEK using the FGD system.  A more serious background
comes from neutral current (NC) interactions where a single $\pi^0$ is
produced and one gamma-ray from its decay is not reconstructed.  The NC
$\pi^0$ production rate was measured using the 1KT, constraining the cross
section.
%%%

%%% Selection %%%
\begin{figure}
  \centering
  \includegraphics[width=8.0cm]{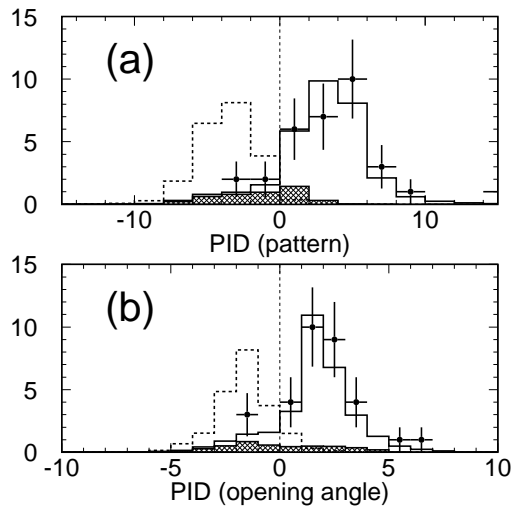}
  \caption{
    Distributions of PID parameters for 32 single-ring
    events based on (a) Cherenkov ring pattern, and (b) Cherenkov
    opening angle.
    The distributions for the data (closed circles),
    oscillated $\nu_\mu$ MC with
    ${\rm (sin^22\theta_{\mu\mu},\Delta m^2_{\mu\mu})}$ $=$ ${\rm (1.0,2.8\times10^{-3}eV^2)}$
    (solid histograms)
    and expected $\nu_e$ signal with full mixing (dashed histograms)
    are shown.
    Shaded histograms are the NC component of $\nu_\mu$ MC.
    }
  \label{fig:pid}
\end{figure}

This analysis is based on the data taken between June 1999 and July 2001,
corresponding to ${\rm 4.8\times 10^{19}}$ protons on target (POT).  A
total of 56 fully-contained events are obtained in the 22.5~kt fiducial
volume of SK.  Single-ring events are selected to enhance CC-QE
interactions against the $\pi^0$ production background.  The details of the
selection criteria for fully-contained single-ring events are found in
Ref.~\cite{K2Kosci:2002up}.  Particle identification (PID) is applied to
reduce $\nu_\mu$-induced backgrounds.  A Cherenkov ring produced by an
electron is diffused by its electromagnetic shower and multiple scattering,
while that produced by a muon has a clear edge, and a low energy muon has a
smaller opening angle than an electron.  Both the Cherenkov ring pattern
and opening angle are required to be consistent with an electron event.
PID parameters are calculated from a log-likelihood difference for the
electron and the muon hypothesis.  Distributions of these PID parameters
are shown in Fig.\ref{fig:pid}.  Negative values of the parameters indicate
an electron-like event.  Distributions of data are consistent with the
oscillated $\nu_\mu$ MC
with ${\rm (sin^22\theta_{\mu\mu},\Delta m^2_{\mu\mu})=(1.0,2.8\times10^{-3}eV^2)}$,
the best-fit parameters of the \numu disappearance analysis in K2K~\cite{K2Kosci:2002up},
where $\theta_{\mu\mu}$ is the effective mixing angle in 2-flavor approximation for $\nu_\mu \rightarrow \nu_x$ oscillation.
The visible energy is
required to be larger than 100~MeV to reject low momentum charged pions and
electrons from muon decays in which muons are below Cherenkov threshold.
Finally, events which are followed by a decay electron signal within a
30~$\mu$sec time window are rejected.  A small fraction of \nue
interactions are also rejected by this cut when they are accompanied with
decay electrons originating from pions in inelastic interactions.  The
overall efficiency to select CC interactions from the oscillated $\nu_e$ is
57~\% for $\Delta m^2=2.8\times 10^{-3}$eV$^2$.

One event is selected as an electron candidate as summarized in
Table~\ref{tbl:reduction}.  While ring-counting algorithms evaluate this
event as a single-ring event, under a careful manual examination it reveals
that the remaining PMT hits out of the reconstructed ring form an
additional ring, and the invariant mass of these two rings is found to be
consistent with the $\pi^0$ mass.  Thus, we conclude that the observed
event is consistent with a $\pi^0$ background event from NC interaction.

\begin{table}
 \begin{center}
 \caption{
    Summary of the event reduction for \nue appearance search at SK. The first
 column lists each reduction step, and the second gives the number of observed
 events after each selection. The numbers of expected background
 from \numu and beam \nue without neutrino oscillations are shown in the
 third and fourth column, respectively. The last column corresponds to the
 expected number of CC interaction events induced by $\nu_e$ oscillated from
 $\nu_\mu$ with ${\rm (sin^22\theta_{\mu e},\Delta m^2_{\mu e})=(1.0,2.8\times10^{-3}eV^2)}$.
 }
 \label{tbl:reduction}
 \begin{tabular}{l|c|c|c|c}
  \hline\hline
 & DATA & $\nu_\mu$ w/o osc & beam $\nu_e$ & $\nu_e$ from \numu osc\\
  \hline
%
% generated & & 104 events & 1.0 events & 28 events \\
%  \hline
%
 FCFV      & 56 & 80 & 0.8 & 28 \\
 Single Ring & 32 & 50 & 0.5 & 20 \\
 PID (e-like) & 1 & 2.9 & 0.4 & 18 \\
 E$_{vis}>$100MeV & 1 & 2.6 & 0.4 & 18 \\
 w/o decay-e & 1 & 2.0 & 0.4 & 16 \\
  \hline\hline

 \end{tabular}
 \end{center}
\end{table}
%%%

%%% Background %%%
In our simulation to predict the number of background events, NC and CC
pion production are modeled following Rein and Sehgal~\cite{Singlepi} for
the resonance region, and GRV94~\cite{Bluck:1995uf} with the correction of
Bodek and Yang~\cite{Bodek:2002md} for the deep inelastic scattering
region.  The axial vector mass for QE and resonance pion production in the
simulation are 1.1~GeV/c$^2$ and 1.2~GeV/c$^2$, respectively.  The 
interaction models used in this analysis are the same as in the \numu{}
disappearance analysis in Ref.~\cite{K2Kosci:2002up} except for the
normalization of NC with respect to CC-QE cross-section.  The reduction of
the background events is summarized in Table~\ref{tbl:reduction}.

\begin{table}
\begin{center}
  \caption{Systematic errors in the expected number of $\nu_\mu$ background
    in SK.} 
  \label{tbl:sksyssummary}
\begin{tabular}{lcc}
\hline\hline
                        & Jun.1999 & Nov.1999$\sim$Jul.2001 \\
\hline
  NC cross-section      & $^{+22\%}_{-27\%}$  &  $^{+20\%}_{-25\%}$ \\
  Ring Counting         & $^{+15\%}_{-13\%}$  &  $^{+15\%}_{-13\%}$ \\
  Particle ID           & $ \pm 11\% $        &  $ \pm 11\% $ \\
  \numu Energy Spectrum & $ \pm 14\% $        &  $ \pm 1\% $ \\
  Far/Near ratio        & $^{+15\%}_{-11\%}$  &  $ \pm 6\%$ \\
$\epsilon_{\mathrm 1KT}$ & $ \pm 4\%$        & $ \pm 4\%$  \\
$\epsilon_{\mathrm SK}$ & $ \pm 3\%$        & $ \pm 3\%$ \\
POT normalization       & $ \pm 0.9\%$        & $ \pm 0.6\%$ \\
CC-nQE cross-section    & $ \pm 1\% $         & $ \pm 0.4\% $ \\
\hline
Total                   & $\pm 36\% $         & $^{+33\%}_{-31\%}$ \\
\hline\hline
\end{tabular}
\end{center}
\end{table}

The expected background from $\nu_\mu$ interactions in the case of no
oscillation is estimated to be 2.0 events where the normalization is
determined by extrapolating the observed number of events in the
1KT~\cite{K2Kosci:2002up}.  It is estimated to be 1.9 events in the case of
$\numu \rightarrow \nutau$ oscillation with ${\rm (sin^22\theta,\Delta
  m^2)=(1.0,2.8\times10^{-3}eV^2)}$.  Since the background is dominated by
NC $\pi^0$ production (87\%) and the oscillated \nutau has the same NC
interactions, it is insensitive to the \numu disappearance oscillation
parameters.

The NC $\pi^0$ production cross-section in the MC simulation is checked by
a 1KT measurement of $\pi^0$ events.  In the 1KT, $\pi^0$ events are
selected by requiring two e-like rings whose invariant mass is between
85~MeV and 215~MeV.  Muon events are also collected by requiring a single
$\mu$-like ring in the 1KT as a reference to \numu flux.  The NC/CC-QE
cross section ratio is calculated from the ratio of the number of $\pi^0$
events to that of muon events.  The $\pi^0$ sample in the 1KT is dominated
by NC interactions (87\%), while the muon sample is dominated by CC
interactions (97\%).  The measured NC/CC-QE ratio is consistent with the MC
simulation; the ratios of data to MC is $1.07^{+0.20}_{-0.15}$.  To cover
the allowed range, from 0.92 to 1.27, without changing the NC cross-section
model in our MC, an uncertainty of 30\% is assigned on the NC/CC-QE ratio.
This error is used to estimate the systematic uncertainty in the \numu
background.
The uncertainty in the cross-section ratio of CC interactions other than QE (CC-nQE) to CC-QE is estimated to be $\pm 20\%$ as in Ref.~\cite{K2Kosci:2002up}

The uncertainty in the $\nu_\mu$-induced background is estimated to be $\pm
0.6$ events for the 2.0 events.  Contributions from various sources to the
systematic error are summarized in Table~\ref{tbl:sksyssummary}.  Since the
horn current and target diameter were different in June 1999 from the other
period, systematic errors are estimated separately for these two periods
and properly weighted to obtain the total systematic error.  The
uncertainty in the NC cross-section gives the largest contribution of
$^{+20}_{-25}\%$.  Systematic errors from ring counting and PID are
estimated by comparing the shape of the MC and data likelihood
distributions for cosmic-ray muons and atmospheric neutrino events.  They
are assigned to be $^{+15}_{-13}$\% and $\pm 11$\%, respectively.
Systematic errors from the neutrino energy spectrum ($\pm 1.0\%$) and
far/near ratio ($\pm 6.0\%$) are estimated in the same manner as in
Ref.~\cite{K2Kosci:2002up}.  Systematic errors from the fiducial volume
definition and detection threshold in the 1KT~($\epsilon_{\mathrm 1KT}$)
and SK~($\epsilon_{\mathrm SK}$) are estimated to be $\pm 4\%$ and $\pm
3\%$, respectively.

The expected background from beam \nue{} interactions in SK is estimated to
be 0.4 events, which is derived from the \nue{}/\numu{} flux ratio
predicted by the beam MC simulation and the \numu flux extrapolated from
the 1KT measurement.  The systematic uncertainty in the number of beam
$\nu_e$ events is estimated to be 0.11 events, which is dominated by the
uncertainty in the $\nu_e$ energy spectrum.  The $\nu_e/\nu_\mu$ ratio has
been verified by a measurement of $\nu_e$ events in the FGD~\cite{Yoshida}.
The $\nu_e$ events in the FGD are selected by requiring 1) a vertex inside
the SciFi fiducial volume, 2) an energy deposit in the PSH of greater than
20~MeV, 2.5 times larger than expected from a muon, 3) an energy deposited
in the LG of greater than 1~GeV, and 4) no corresponding hits in the MRD.
During an exposure of $2.9\times 10^{19}$~POT, 51 electron candidates are
selected with an estimated background of 24 \numu{} induced events.  The
$\nu_e/\nu_\mu$ interaction ratio is measured to be $1.6 \pm 0.4(stat.)
^{+0.8}_{-0.6}(sys.)$\% which is in agreement with the beam MC prediction
of 1.3\%.
%%%

%%% Oscillation %%%
The observation of one electron event in SK is consistent with the expected
background of 2.4 events in the case of no oscillation.
A constraint on neutrino oscillations from \numu to \nue is obtained by
comparing the observed number of electron events with the expectation assuming
oscillations.
The expected number of electron events is calculated by
\begin{equation}
  N_{exp} =  N^{OSC}_{\nu_e} + N^{BG}_{\nu_\mu} + N^{BG}_{\nu_e},
  \label{eq:numberofelectronevents}
\end{equation}
where $N^{OSC}_{\nu_e}$ is the number of electron events induced by
oscillated \nue, $N^{BG}_{\nu_e}$ is that induced by beam $\nu_e$, and
$N^{BG}_{\nu_\mu}$ is that induced by both CC and NC interactions of
$\nu_\mu$ and NC interactions of \nue and \nutau from oscillations.  The
$\numu \rightarrow \nue$ oscillation signal, $N^{OSC}_{\nu_e}$, depends on
the probability of \nue appearance expressed by Eq.~\ref{eq:appearance}.
The number of beam $\nu_e$
induced background, $N^{BG}_{\nu_e}=0.4$, is treated as a constant, since a
contribution of $\nue \rightarrow \nux$ oscillation is negligible.  The CC
component of $N^{BG}_{\nu_\mu}$ decreases with $\nu_\mu$ disappearance
observed in K2K and atmospheric neutrino experiments , depending on the
survival probability,
\begin{equation}
  P(\mumu{})
  = 1-\sin^2 2\thetamumu{}  
  \sin^2(1.27\dmsmumu{} L/E),
  \label{eq:disappearance}
\end{equation}
where \thetamumu{} and \dmsmumu{} are the effective mixing angle and mass
squared difference for \numu{} disappearance, respectively.  In the present
analysis, we assume $\thetamumu{} = \frac{\pi}{4}$ based on the nearly full
mixing observed by atmospheric neutrino experiments, and $\Delta
m_{\mu\mu}^2 = \Delta m_{\mu e}^2$, which is implied in the framework of
3-flavor neutrino mixing by the small mass difference found in solar
neutrino experiments~\cite{PDBook}.  Thus, $N_{exp}$ reduces to a
function of two parameters, $\theta_{\mu e}$ and $\Delta m_{\mu e}^2$.

A probability density function~(PDF) for $N_{exp}$ is constructed from the
Poisson distribution convoluted with the systematic uncertainty.  Given the
observation of one electron event, the systematic uncertainty has a very
small effect on the derived confidence interval.
\begin{figure}
  \includegraphics[width=8cm]{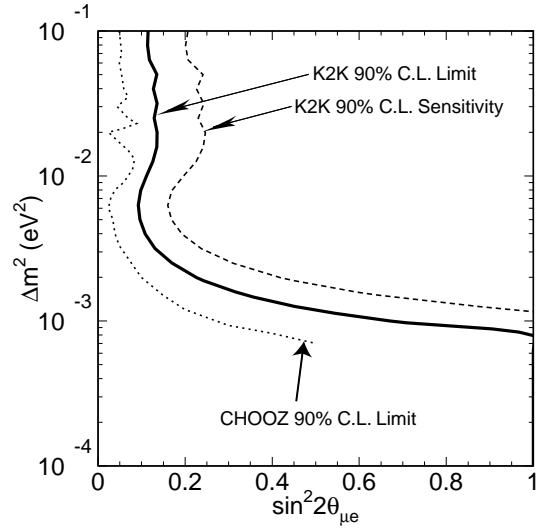}
  \caption{
    \label{fig:contour}
    The confidence interval for \mue{} oscillations as a function of the
    effective \dmsmue{} at 90\% C.L. (solid line).
    Dashed line indicates 90\% C.L. sensitivity of the experiment
    for the current statistics.
    The area to the right of each curve is excluded.
    Dotted line shows the limit at 90\% C.L. by CHOOZ
    assuming $\sin^2 2 \theta_{\mu e} = \frac{1}{2} \sin^2 2 \theta_{13}$.}
\end{figure}
The confidence interval of \ssttmue{} is calculated using the method
suggested in Ref.~\cite{Feldman:1998qc}.  In the calculation, the best-fit
parameters are searched for in the 2-dimensional parameter space with
\ssttmue{} bounded in [0,1].

Figure~\ref{fig:contour} shows the limit on \ssttmue{} as a function of
\dmsmue{}.  The experimental limits on the neutrino mixing for the $\numu
\rightarrow \nue$ oscillation hypothesis are given at 90\% C.L. for a
parameter region with $\Delta m_{\mu e}^2 > 6\times10^{-4}$~eV$^2$.
Neutrino oscillations from \numu to \nue are excluded at 90\% C.L. for
$\ssttmue{} > 0.15$ at \dmsmue{} $ = 2.8\times 10^{-3}$eV$^2$.  The most
stringent limit of $\ssttmue{} < 0.09$ is set for $\dmsmue{} =
6\times{}10^{-3}$~eV$^2$.
The sensitivity of the experiment for the current statistics is also shown
in Fig.~\ref{fig:contour}.

Assuming 3-flavor neutrino oscillations and CPT invariance, our results can
be compared to reactor experiments.  CHOOZ has excluded $\sin ^2
2\theta_{13}>0.1$ at $\dmsonethree{} \sim 3\times
10^{-3}$eV$^2$~\cite{CHOOZ}.  This corresponds to a limit of $\ssttmue{} <
0.05$ at $\dmsmue{} \sim 3\times 10^{-3}$eV$^2$ assuming $\theta_{23} =
\pi/4$, and consistent with the present analysis as shown in
Fig.\ref{fig:contour}.  The limit on $\sin^2 2 \theta_{13}$ by CHOOZ is
converted by assuming $\sin^2 2 \theta_{\mu e} = \frac{1}{2} \sin^2 2
\theta_{13}$.
%%%

%%% Conclusions %%%
The K2K experiment searched for $\numu \rightarrow \nue$ oscillations with
accelerator-produced muon neutrinos traveling 250~km. This is the first
experimental search for \nue appearance with sensitivity down to the \dms{}
suggested by atmospheric neutrino oscillations.  A single electron
candidate is found in SK.  The observed event is consistent with the
expected background event.  The limit on the \nue appearance is obtained.
At the best-fit parameter values of the K2K $\nu_\mu$ disappearance
analysis, we set the 90\% confidence limit of $\ssttmue{} < 0.15$.
%%%

%%% Acknowledgments %%%
We thank the KEK and ICRR Directorates for their strong support and
encouragement.  K2K is made possible by the inventiveness and the
diligent efforts of the KEK-PS machine and beam channel groups.  
We thankfully appreciate discussions on the statistical treatment
with Dr. Louis Lyons.
We gratefully
acknowledge the cooperation of the Kamioka Mining and Smelting
Company.  This work has been supported by the Ministry of Education,
Culture, Sports, Science and Technology, Government of Japan and its
grants for Scientific Research, the Japan Society for Promotion of
Science, the U.S. Department of Energy, the Korea Research
Foundation, the Korea Science and Engineering Foundation,
the CHEP in Korea, and Polish KBN grant  5P03B06531.
%%%

%%% References %%%

%%%
\end{document}